\documentclass[fleqn,10pt]{wlscirep}
\usepackage[utf8]{inputenc}
\usepackage[T1]{fontenc}
\usepackage{bm}
\title{Mechanobiology of Collective Cell Migration in 3D Microenvironments}

\author[1]{Alex M. Hruska}
\author[2]{Haiqian Yang}
\author[3]{Susan E. Leggett}
\author[2]{Ming Guo}
\author[1,*]{Ian Y. Wong}
\affil[1]{School of Engineering, Center for Biomedical Engineering, Joint Program in Cancer Biology. Brown University, Providence RI 02912.}
\affil[2]{Department of Mechanical Engineering, Massachusetts Institute of Technology, Cambridge, MA 02139.}
\affil[3]{Department of Chemical and Biological Engineering, Princeton University, Princeton, NJ 08540.}
\affil[*]{E-mail: ian\_wong@brown.edu}

\begin{abstract}
Tumor cells invade individually or in groups, mediated by mechanical interactions between cells and their surrounding matrix. These multicellular dynamics are reminiscent of leader-follower coordination and epithelial-mesenchymal transitions (EMT) in tissue development, which may occur via dysregulation of associated molecular or physical mechanisms. However, it remains challenging to elucidate such phenotypic heterogeneity and plasticity without precision measurements of single cell behavior. The convergence of technological developments in live cell imaging, biophysical measurements, and 3D biomaterials are highly promising to reveal how tumor cells cooperate in aberrant microenvironments. Here, we highlight new results in collective migration from the perspective of cancer biology and bioengineering. First, we review the biology of collective cell migration. Next, we consider physics-inspired analyses based on order parameters and phase transitions. Further, we examine the interplay of metabolism and heterogeneity in collective migration. We then review the extracellular matrix, and new modalities for mechanical characterization of 3D biomaterials. We also explore epithelial-mesenchymal plasticity and implications for tumor progression. Finally, we speculate on future directions for integrating mechanobiology and cancer cell biology to elucidate collective migration.
\end{abstract}
\begin{document}

\flushbottom
\maketitle

\thispagestyle{empty}

\section{Introduction}
\label{sec:intro}
Collective migration occurs when a group of individual cells exhibit coordinated movements with similar speed and directionality \cite{Rorth2009}. This phenomenon is intriguing in cancer biology as a demonstration of how heterogeneous phenotypes cooperate to enhance solid tumor progression within a dysregulated microenvironment \cite{Tabassum2015}. Indeed, patient histopathology sections often include packs of tightly connected cancer cells organized as rounded clusters or linearly extended strands within an aberrant tissue architecture \cite{Friedl2012ncb}. Such behaviors are reminiscent of leader-follower interactions in development and wound healing \cite{VilchezMercedes2021}. Moreover, the epithelial-mesenchymal transition (EMT) is associated with developmental transcription factors that weaken cell-cell adhesions and promote tumor invasion \cite{Yang2020b}. Collective migration is also intriguing from a physical science perspective as an emergent phenomenon driven by interactions between many individuals \cite{Trepat2018}. These behaviors are further influenced by abnormal solid and fluid stresses from their surroundings, which impact metastatic dissemination and drug delivery \cite{Jain2020}. Thus, a cross-disciplinary approach is required to understand how heterogeneous groups of invasive cells interact with the surrounding matrix via biological and physical mechanisms.

Recent advances in assay development have enabled new mechanistic insights into collective migration via exquisitely sensitive measurements as well as biomimetic culture conditions. For example, single cell molecular profiling technologies have revealed the interplay of myriad signaling pathways, which previously involved dissociation of tissues into single cells \cite{Satija2019}, but now can retain information about tissue architecture via spatial transcriptomics \cite{rao2021exploring}. New light microscopy techniques based on structured illumination permit deeper imaging with improved temporal resolution and reduced phototoxicity \cite{Lemon2020}. Fluorescent protein reporters can be used for longitudinal monitoring of transcriptional state or force transduction \cite{Specht2017}. In parallel, cells cultured within engineered 3D biomaterials experience controlled biochemical and physical cues, which can be dynamically modulated \cite{Leggett2017,Beri2018}. The mechanical properties of single cells and local matrix architecture can be further probed at subcellular resolution \cite{Leggett2021, LI20211863}. These early stage technologies are now being translated with primary mouse or human organoids in order to predict disease progression and therapeutic response \cite{LeSavage2021}.  

In this chapter, we highlight recent developments in collective tumor invasion from a biological and physical perspective. We first review the essential concepts of collective cell migration and leader-follower interactions. We then examine how patterns of collective motion can be analyzed using physical concepts such as order parameters and phase transitions, as well as the interplay of metabolism with phenotypic heterogeneity. Next, we review biomaterials and the ECM, along with precision measurement 3D matrix mechanics. Finally, we explore EMT as a representation of phenotypic heterogeneity, including recent results on partial or intermediate states.  We close with our critical perspective on the field and discussion of future directions.

\section{Collective Cell Migration and Leader-Follower Interactions}
\label{sec:leader}

Early work on eukaryotic cell migration investigated how individual cells (e.g. fibroblasts, leukocytes) cyclically adhere and propel themselves along planar (2D) substrates, revealing physically conserved mechanisms across various cell types \cite{lauffenburger1996cell}. First, cells polarize with a leading and trailing edge, often in response to asymmetric stimuli of soluble or immobilized biochemical factors, substrate stiffness, or local topography \cite{sengupta2021principles}. Next, cells extend protrusions from the leading edge driven by actin polymerization, which form integrin-mediated focal adhesions with ligands on the substrate \cite{ridley2003cell}. Finally, actomyosin contractility via stress fibers pulls the cell body forward and retract the trailing edge \cite{levayer2012biomechanical}. Subsequent work using compliant substrates revealed that cell adhesion and migration were modulated by the relative strength of cell-substrate interactions and intracellular contractility \cite{Janmey2020}.

In comparison, 3D matrix is a more mechanically confined environment that can impede cell migration relative to 2D substrates \cite{yamada2019mechanisms}. Migratory cells can traverse a narrow region by local remodeling of the matrix into a wider path, as well as by deforming their intracellular components  to squeeze through. Thus, the increased deformability reported for cancer cells relative to normal cells may enhance invasion and metastatic dissemination \cite{Alibert2017}. Classically, individual migration phenotypes have been described in terms of contractile mesenchymal or propulsive amoeboid modes. First, mesenchymal migration phenotypes are associated with elongated, spindle-like morphologies, which extend narrow actin-driven protrusions (via RAC1 and CDC42) that apply strong cell-matrix adhesions and matrix metalloproteinases to degrade and remodel the ECM \cite{yilmaz2009emt}. Mesenchymal cells express the intermediate filament vimentin \cite{Leggett2021}, which is highly stretchable (without breaking) and can protect cells undergoing large deformations \cite{Hu2019, Patteson2019}. Second, amoeboid migration phenotypes exhibit compact, rounded morphologies, with propel themselves forward via actomyosin contractility (via RHOA) and do not require strong adhesions or matrix metalloproteinases for effective migration \cite{paluch2016focal}. Weak ECM adhesion and limited remodeling allow cells utilizing amoeboid migration (e.g. immune cells) to migrate more rapidly than mesenchymal cells (e.g. fibroblasts) which tend to migrate more slowly \cite{Liu2015, Ruprecht2015}. Migratory cells are further capable of switching between mesenchymal and amoeboid migration modes, depending on local ECM adhesivity and matrix remodeling capabilities \cite{Wolf2003a, Sahai2003}. Further, groups of mesenchymal or amoeboid cells may migrate as multicellular ``streams'' without strong cell-cell adhesions in response to some asymmetric external stimulus, where contact guidance by matrix architecture is permissive for similar migration direction and speed \cite{roussos2011chemotaxis}. In the absence of these asymmetric stimuli, cells typically exhibit more random and less directed motility \cite{sengupta2021principles}. Interestingly, an osmotic-pressure driven propulsion mechanism has also been observed for cancer cell lines in confining microchannels, which does not require actin polymerization \cite{Stroka2014a}.

Within cancer cells, the nucleus is the most rigid and sizable organelle, undergoing large deformations when translocating through confined spaces \cite{Kalukula2022}. Empirically, cancer cell lines cannot traverse confined spaces smaller than 7 $\mu$m$^2$ without matrix remodeling \cite{wolf2013physical,Davidson2014}, which is gauged by nuclear deformation as an intracellular ``ruler'' \cite{Renkawitz2019,Lomakin2020}. In particular, nuclear stiffness is mediated in part by the nuclear envelope via nucleoskeletal protein lamin A/C \cite{Harada2014}, and lamin A has been reported to be downregulated in breast cancer cells \cite{Bell2021.07.12.451842}. The nucleus is mechanically coupled to the cytoskeleton via the linker of nucleoskeleton and cytoskeleton (LINC) complexes, allowing nuclear pushing or pulling via actomyosin contractility \cite{Kalukula2022}. As a consequence of these large deformations, nuclear envelope ruptures can occur and drive DNA damage, which may only be partially reversed by ESCRT III \cite{Denais2016,Raab2016,Irianto2017a,Xia2018,Shah2021}, and may enhance metastatic dissemination \cite{Fanfone2022}. Indeed, cells can even use the nucleus as a ``piston'' to compartmentalize the front and back to generate high pressure ``lobopodial'' protrusions within very dense matrix architectures \cite{Petrie2014,Petrie2017}.

\begin{figure}[h!]
\includegraphics[width=4.5in]{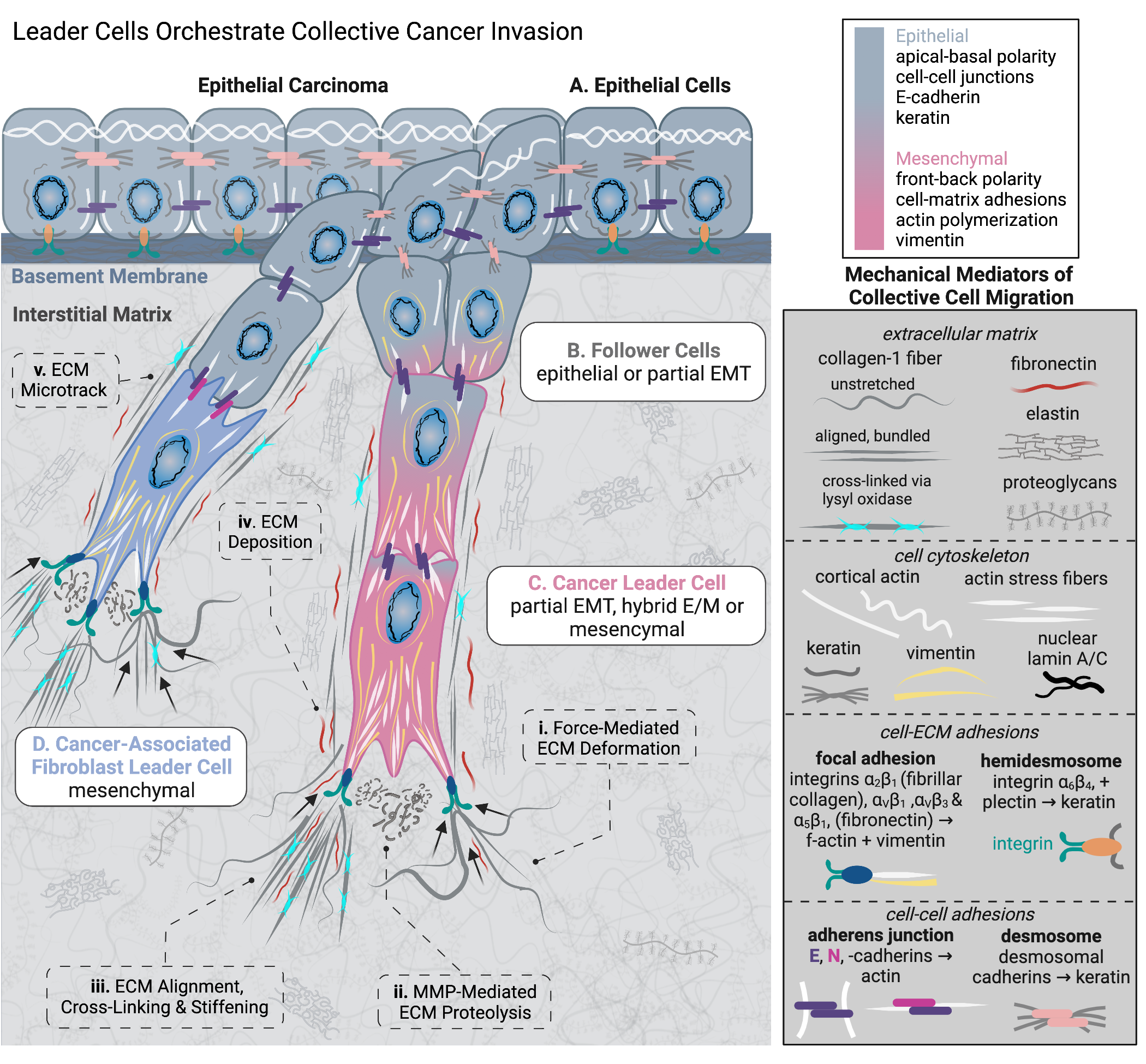}
\caption{Schematic of collective cancer cell invasion. A. Epithelial cells, B. Follower cells, C. Cancer Leader Cell, D. Cancer-Associated Fibroblast. ECM remodeling via force-mediated matrix deformations (i), MMP degradation (ii), alignment / crosslinking (iii), ECM deposition (iv), and formation of ECM microtracks (v).}
\label{fig:leader_cells}       
\end{figure}

Carcinomas arise from healthy epithelial tissues, which consist of epithelial cells with strong cell-cell adhesion and apicobasal polarity organized into tightly packed layers (Fig. \ref{fig:leader_cells}A) \cite{rodriguez1989morphogenesis}. During tumor progression, tissues become progressively disorganized due to uncontrolled proliferation, then transition to malignant invasion via degradation of the basement membrane \cite{Kelley2014a}.  Groups of epithelial cells exhibit collective migration when they remain mechanically connected, sustaining cell polarization for directed motion, as well as coordinating longer ranged responses to environmental stimuli (Fig. \ref{fig:leader_cells}B,C) \cite{Rorth2009}. Within solid tumors, the spatial organization of these groups can vary widely, from relatively compact clusters to elongated single or multi-file strands to large masses \cite{Friedl2012ncb}. These collectives may be categorized broadly into distinct functional groups based on parameters of multicellular morphology, the degree of cell-cell adhesion, and supracellular coupling of intercellular signaling  \cite{mayor2016front}. Furthermore, collective migration patterns are shaped (in part) by the surrounding ECM architecture and presence of various stromal cells. 

Despite differences in spatial organization, 3D collective invasion is most often characterized by the presence of highly motile cells at the front of invasive groups, termed ``leader cells'' (Fig. \ref{fig:leader_cells}C), \cite{VilchezMercedes2021}. Leader cells 
often exhibit actin-rich protrusions with strong cell-matrix adhesions and actomyosin contractility to locally remodel the ECM (Fig. \ref{fig:leader_cells}C), analogous to the mesenchymal motility phenotype described previously \cite{mayor2016front}. ECM adhesion is mediated via heterodimeric surface receptors called integrins, such as $\alpha_2\beta_1$ for fibrillar collagen, as well as $\alpha_V\beta_1$, $\alpha_V\beta_3$, and $\alpha_5\beta_1$ for fibronectin, although integrins exhibit some promiscuity and context-dependent function \cite{Hamidi2018}. Leader cells further generate traction forces to deform the ECM via contractility of intracellular stress fibers, which can associate with vimentin intermediate filaments in order to enhance tractions \cite{Leggett2021} (Fig. \ref{fig:leader_cells},i), which can occur in combination with ECM degradation via matrix metalloproteinases (Fig. \ref{fig:leader_cells},ii). Thus, through repeated protrusion-contraction-deformation events, leader cells pull on intrinsically ``wavy'' collagen fibers to stretch and bundle fibers into regions of aligned, stiffened ECM (Fig. \ref{fig:leader_cells},iii). Furthermore, collagen fibers can be cross-linked enzymatically by lysyl oxidase to locally enhance ECM stiffness and promote tumor invasion though focal adhesion kinase (FAK) signaling \cite{Levental2009}. Finally, leader cells can deposit new ECM (Fig. \ref{fig:leader_cells},iv), such as fibronectin, to reinforce adhesive migratory cues \cite{Winkler2020}. Altogether, these leader cell-driven ECM remodeling activities result in the generation of ECM "microtracks" (Fig. \ref{fig:leader_cells},v) which are permissive for cell migration and contain aligned, bundled ECM proteins that present contact guidance cues. For the remainder of this review, we primarily focus on leader cells originating from epithelial tumors, or carcinomas, but note that leader cells may also arise from stromal cells (e.g. fibroblasts) (Fig. \ref{fig:leader_cells}D) or immune cells (e.g. macrophages) \cite{VilchezMercedes2021}. Regardless of cell type, leader cell activity typically results in ECM architecture that is more permissive for cell migration.

Concomitant with ECM remodeling activities, leader cells may guide the migration of adjacent cells termed "followers"  (Fig. \ref{fig:leader_cells}B) along ECM microtracks. Follower cells are defined primarily by a rearward spatial position relative to leaders but may also possess migratory and ECM remodeling capabilities. Indeed, follower cells traversing wider spaces generated by leader cells encounter less mechanical resistance and can expend less energy to remodel ECM or undergo large deformations \cite{VilchezMercedes2021}. Leader cells coordinate the activity of followers through a combination of cell-cell mechanical adhesions and biochemical signaling. 

First, cell-cell junctions are formed when two cells connect adhesion proteins to form a stable mechanical junction (e.g. adherens junctions and desmosomes), typically utilizing proteins in the cadherin superfamily (e.g. E, N, P-cadherins or desmosomal cadherins) \cite{cavallaro2004cell}, which are intracellularly coupled to actin microfilaments or keratin intermediate filaments, respectively \cite{jacob2018types}. As a stationary sheet-like structure, epithelial cells remain tightly-connected to neighboring cells via desmosome adhesions and adhere to the basement membrane via hemidesmosomes containing integrin $\alpha_6\beta_4$ (Fig. \ref{fig:leader_cells}A). Loss of apical-basal polarity is a common step in cancer progression, which often accompanies loss of cell-cell adhesions and acquisition of cell-ECM adhesions, which establish front-back polarity and induce cell motility (Fig. \ref{fig:leader_cells}B). Motile follower cells may lose desmosomes and keratin expression and gain expression of vimentin IF's, which are associated with EMT. However, hybrid phenotypes expressing keratin and vimentin have been observed clinically \cite{Thomas1999}. Mesenchymal leader and followers may remain adherent through N-cadherin junctions, or maintain transient adhesions that result in weak cell-cell coordination. Nonetheless, epithelial followers often retain E-cadherin to tightly coordinate migration. Finally, leader cells are connected to followers via adherens junctions. For example, cancer leader cells in partial EMT states may connect to followers via homotypic E-cadherin junctions (Fig. \ref{fig:leader_cells}C). Leaders in later stages of EMT may express N-cadherin and adhere to followers via heterotypic E/N-cadherin junctions, which have also been documented for fibroblast leader cells with epithelial follower cells (Fig. \ref{fig:leader_cells}D) \cite{Labernadie2017}. Overall, cell-cell adhesions maintain coherent collective motion by mechanical coupling via the cell cytoskeleton, as well as reinforcement of multicellular front-rear polarity. From a biochemical standpoint, cells in close proximity to one another may also coordinate motility by juxtacrine signaling (e.g. Notch1-Dll4 signaling) \cite{Riahi2015}. Moreover, contact-independent autocrine signaling across gap junctions may be utilized for intercellular coordination of contractility and size \cite{Aasen2019}, via connexins \cite{Han2020a, khalil2020collective}.

It should be noted that leader cells may not always guide collective migration, as exemplified by the diversity of collective migration behaviors in development, such as the attraction-repulsion dynamics observed in neural crest-placode migration \cite{Theveneau2013a}. Furthermore, followers may or may not have the potential to become leader cells. Interestingly, followers can ``switch'' with leaders during collective migration, which may occur if leader cells undergo cell division, and could also redistribute metabolic costs of migration. Nevertheless, the functional role and molecular definition of leader cells remains unresolved, and may depend on biological context \cite{theveneau2017leaders}. In this chapter, we consider leader cell behavior and leader-follower dynamics in the context of epithelial cancer invasion. As discussed previously, leader cells often utilize cell-cell adhesions through proteins such as E-cadherin, which is an epithelial biomarker \cite{Leggett2019}. Thus, these leader cells may represent a hybrid or partial EMT phenotype, which we consider in detail in a subsequent section. 

\section{Analyzing Collective Migration using Phase Transitions and Order Parameters}
\label{sec:order}

\begin{figure}[h]
\includegraphics[width=4.5in]{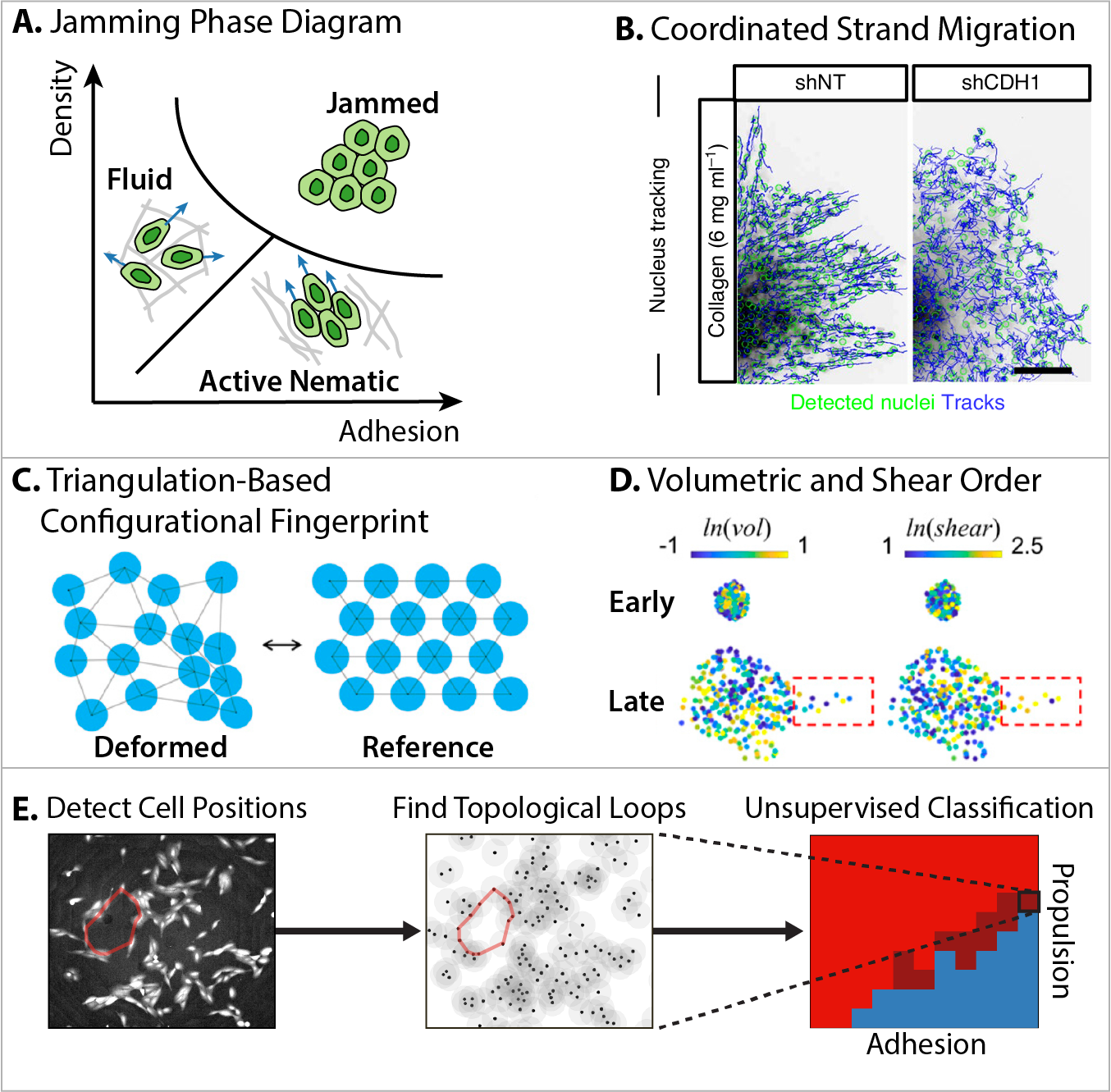}
\caption{A. ``Jamming'' phase diagram of individual and collective cell migration governed by cell-cell adhesion and cell density. Redrawn from \cite{Porta2020}. B. Representative single cell migration tracks (blue lines) from multicellular spheroids in low density collagen matrix, revealing coordinated and persistent motion for vehicle control (shNT), which is lost after knockdown of E-cadherin cell-cell junctions (sdCDH1). Reproduced from \cite{Ilina2020}. C. Triangulation-based configurational fingerprints compare deformed and (relaxed) reference states. D. Late (invasive) spheroids exhibit increased volumetric order parameter (per cell) relative to early spheroids, but comparable shear order parameters (per cell). Reproduced from \cite{Yang2021}. E. Topological data analysis represents discrete cell positions based on the ``persistence'' of topological loops at varying length scales around empty areas, which can then be mapped to computational simulations with varying adhesion and propulsion. Reproduced from \cite{Bhaskar2021}.}
\label{fig:jamming}       
\end{figure}

Healthy epithelial tissues are comprised of mechanically connected cells with limited motility, as well as uniform shape and regular geometric packing \cite{Guillot2013}. This mechanically arrested state can transition to more motile states during tumor progression, development, and wound healing \cite{Oswald2017}. For example, cells can disperse individually (e.g. gas-like), can migrate in closer proximity (e.g. fluid-like), or move in coordinated groups (e.g. active nematics with long-ranged orientational order) (Fig. \ref{fig:jamming}A) \cite{Porta2020,fredberg2021origin}. In these latter contexts, cells may exhibit varying cell-cell adhesion strength along with larger fluctuations in cell shape and packing. Analogous phase transitions that emerge from many interacting particles can be captured using statistical physics, assuming that the particles are relatively similar, move with constant velocity and can alter direction, interact over some distance, and are subject to ``noise'' of varying strength \cite{Vicsek2012}. These phase transitions can be quantified based on changes in some order parameter that represents a symmetry property of the system \cite{chaikin1995principles}. For instance, the jamming transition describes how soft materials exhibit a liquid to solid-like response when varying control parameters of interparticle adhesion, density, or shear \cite{Liu2010}. Fredberg proposed that a biological jamming transition could occur similarly with respect to cell-cell adhesion, cell density, or motility in 2D epithelial monolayers \cite{Sadati2013}.

An intuitive order parameter for collective migration is cell velocity, and particle image velocimetry (PIV) can be used to analyze spatial and temporal correlations \cite{Adrian2005}. Early work used PIV to visualize cooperative motions within 2D epithelial monolayers. For instance, Silberzan and coworkers utilized PIV for collective migration of epithelial monolayers into unoccupied regions \cite{Poujade2007, Petitjean2010}.  Angelini et al. used PIV to measure how compliant substrates were deformed by epithelial monolayers, revealing characteristic length and time scales over which cells coordinate their migration, analogous to dynamic heterogeneity in a physical glass transition \cite{Angelini2010, Angelini2011}. More recently, Scita and coworkers applied PIV to reveal the onset of angular rotation and local cellular rearrangements in 3D spheroids \cite{Palamidessi2019,Cerbino2021}. 

O. Ilina et al. utilized PIV to establish a physical picture of 3D collective migration based the crosstalk of cell-cell adhesion with ECM confinement \cite{Ilina2020}. Multicellular spheroids were embedded within collagen matrix in varying assay geometries. When ECM porosity was sufficiently large to permit cell migration (without ECM degradation), cells with downregulated E-cadherin typically migrated individually in a fluid-like state. As ECM porosity decreased, the increasing spatial confinement resulted in cells moving together in close proximity, despite their relatively weak cell-cell junctions. Nevertheless, higher resolution tracking of cell velocity and spatial correlations revealed that neighboring cells within a group were locally uncorrelated when cell-cell junctions were weak (Fig. \ref{fig:jamming}B). Thus, these results highlight the importance of mature cell-cell adhesions (e.g. E-cadherin) to coordinate effective group migration during tumor invasion and metastasis.

Geometric order parameters can also be defined based on cell shape or local connectivity. Historically, honeycomb-like hexagon-dominated cellular structures were observed empirically in animal and plant tissues \cite{thompson_1968}. Interestingly, proliferating cells within epithelial tissues result in a wider distribution of polygons, particularly pentagons and heptagons \cite{Gibson2006}. Bi and Manning captured these transitions based on a cell shape index (i.e. ratio of cell perimeter to area), which are indicative of shear modulus of the cell monolayer \cite{Bi2015}, effective diffusivity of single cells \cite{Bi2016} and cellular stresses \cite{Yang2017a}. Fredberg and coworkers subsequently validated this parameter in asthmatic airway epithelium \cite{Park2015}, and revealed that cell shape index and variability were also strongly correlated with unjamming \cite{Atia2018, Mitchel2020}. These concepts were also extended to 3D collective migration by analogy to phase transitions from solid to liquid to gas \cite{Wong2014,Kang2021, Grosser2021}.
 
H. Yang et al. proposed new order parameters based on the deformation of a triangular lattice relative to an idealized equilibrium (Fig. \ref{fig:jamming}C) \cite{Yang2021}. For some spatial configuration of cells within a tissue, the cell positions (e.g. nuclei) can be connected by fictitious lines to partition the surrounding space into triangles (e.g. Delaunay triangulation). The deformation gradient can then be expressed using invariants $vol_n = \det{(\mathbf{F})}$, representing how the area of each triangle deviates from the average, along with $shear_n = \text{tr}\,(\mathbf{F}^T\mathbf{F})/\det{\mathbf{(F)}}-2$, corresponding to how a given triangle is distorted relative to an equilateral triangle. Corresponding order parameters are defined by ensemble averaging the natural logarithm of these invariants over all triangles, i.e. $\Phi_{vol} = \langle [\log{(vol_n)}]^2\rangle$ and $\Phi_{shear} = \langle \log{(shear_n)}\rangle$. This analysis was first validated based on ventral furrow formation of the \emph{Drosophila} germband epithelium, which transitions from an arrested, jammed-like state towards a more motile, unjammed-like state. This macroscopic phenomenon is associated with characteristic shearing and shrinking behavior within the tissue, resulting in a sharp increase in the $\Phi_{shear}$ and $\Phi_{vol}$ at a characteristic timescale where this phase transitions occurs. Conversely, the proliferation of kidney epithelial cells (e.g. MDCK) first results in a suppression of density fluctuations in $\Phi_{vol}$, analogous to a transition from gas to liquid, followed by a further suppression of packing disorder in $\Phi_{shear}$, analogous to a transition from liquid to solid. Lastly, these concepts were used to analyze the collective invasion of mammary epithelial spheroids (e.g. MCF-10A). Remarkably, spheroids at earlier times with minimal invasion (e.g. acini) exhibited similar $\Phi_{shear}$ but smaller $\Phi_{vol}$ relative to spheroids at later time with more invasive strands (Fig. \ref{fig:jamming}D). For these spheroids at later times, cells within these peripheral strands exhibited increased $\Phi_{shear}$ relative to the interior. Although this framework does not require the reference triangles to be the same, and any experimental frames can be used as the reference, special caution should still be taken when choosing appropriate reference frames, especially when dealing with naturally heterogeneous biological systems. Indeed, Y. Han et al. observed that cells within these multicellular spheroids dynamically modulate their size and stiffness as they shuttle between the interior and peripheral strands \cite{Han2020a}. Remarkably, cells within the spheroid core tended to be smaller and stiffer, while cells along invasive strands were appreciably larger and softer. Cells often switched positions between the spheroid core and periphery, and could dynamically regulate their volume and stiffness by fluid exchange through gap junctions. Further investigations will help to establish the validity of this approach across different biological systems.

D. Bhaskar et al. utilized topological data analysis to visualize the ``shape'' of a tissue based on discrete cell positions \cite{Bhaskar2021}. Different spatial configurations of cells can be compared by computing the ``cost'' of rearranging topological features (i.e. persistence homology). Briefly, the spatial connectivity between cell positions is now sampled across varying length scales, generating a topological signature (``barcode'') that emphasizes features that are present across multiple length scales. For pairwise connected components (i.e. dimension 0 homology), this analysis will be skewed by population sizes, making it challenging to compare biological specimens with different numbers of cells. Instead, Bhaskar et al. considered how nuclei can be linked as connected loops around an empty area (i.e. dimension 1 homology), which is less sensitive to population size and also samples larger scale spatial structure. This unsupervised machine learning approach correctly classified experimental data of cell positions in individual, compact clusters, or branching phases (Fig. \ref{fig:jamming}E). Further, this approach could classify distinct phases and identify phase transitions for simulated self-propelled particles with varying motility and adhesion strength. The success of this approach can be explained by mapping discrete data points to a continuous shape (i.e. manifold), which remains highly effective even with relatively sparse sampling, as well as variability in the position of data points. Indeed, different phases could be successfully distinguished even when 80\% of the cells were randomly removed. Ongoing research is extending this approach to multiple cell types.

Order parameters represent a powerful approach to understand non-living physical systems, particularly in soft matter \cite{chaikin1995principles}. From a reductionist perspective, it is appealing to think that these approaches could also describe collective migration \cite{Oswald2017,fredberg2021origin}, although these are living systems that are far from thermodynamic equilibrium. Indeed, groups of cancer cells are likely to be quite heterogeneous and in smaller numbers (than in soft matter systems), and the nature of biological ``noise'' remains poorly understood. For instance, a given population of cancer cells may be comprised of distinct subpopulations with ``epithelial'' or ``mesenchymal'' states, associated with distinct motility and adhesion phenotypes \cite{GamboaCastro2016}. Thus, some care is warranted when using order parameters that average biological behaviors across populations and time. An intriguing possibility is that unsupervised machine learning could be used to classify spatiotemporal patterns of cell migration and infer new order parameters \cite{Cichos2020}. Although so-called deep learning is computationally expensive and typically requires extensive training data, constraining convolutional neural networks with some physical principles (e.g. physics-informed machine learning \cite{Karniadakis2021}) could be highly effective for smaller experimental datasets.

\section{Metabolic Heterogeneity and Mechanobiological Phenotype}
\label{sec:metabolism}
\begin{figure}[h!]
\centering
\includegraphics[width=3.5in]{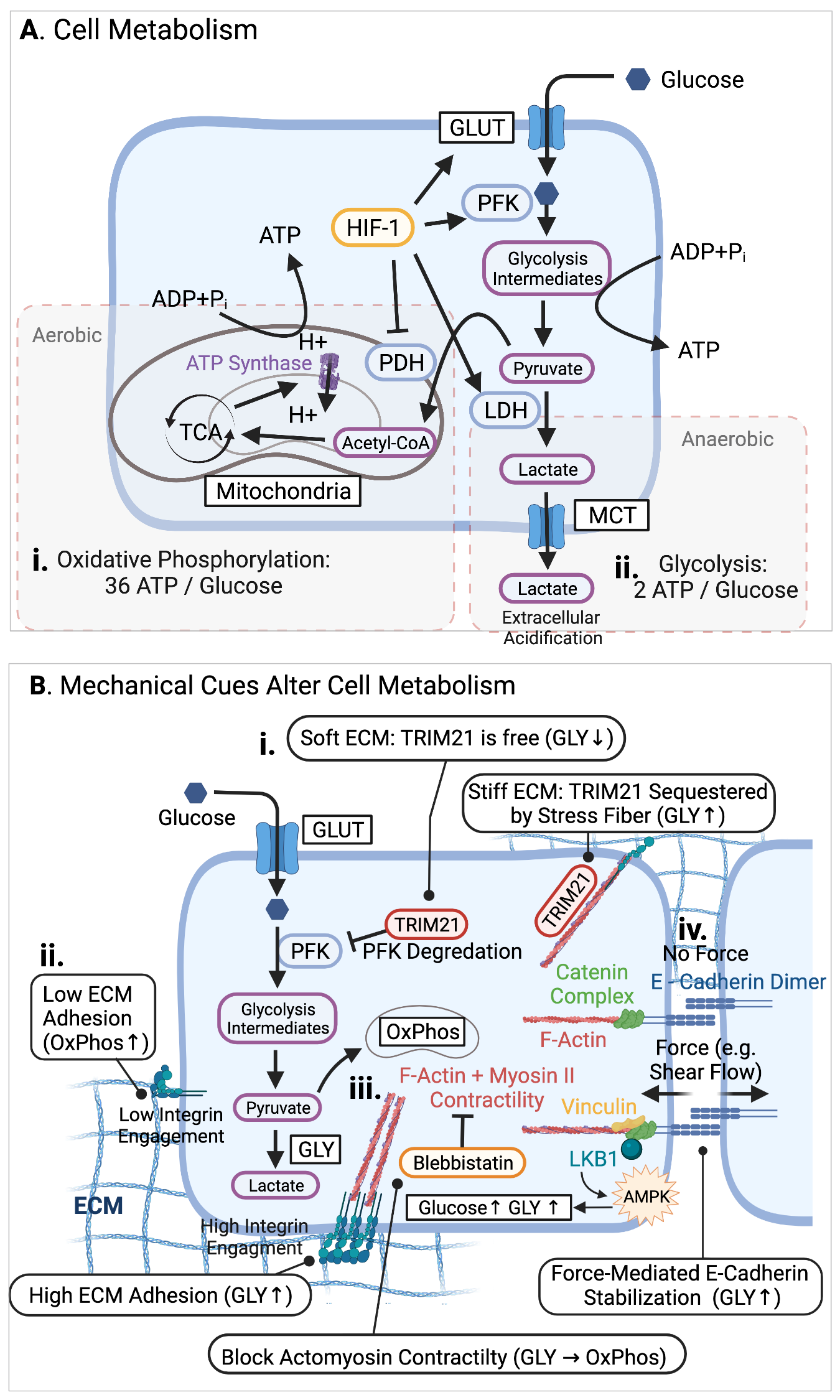}
\caption{A. Cell metabolism converts glucose to ATP via oxidative phosphorylation (i) or glycolysis (ii), depending on oxygen availability. B. Mechanical cues alter cell metabolism via substrate stiffness and TRIM 21 (i), weak cell-matrix adhesion (ii), strong cell-matrix adhesion with actomyosin contractility (iii), or force-mediated E-cadherin stabilization (iv).}
\label{fig:metabolism_background}       
\end{figure}

Tumor cells reprogram their metabolic activity in order to sustain proliferation and invasion in dysregulated microenvironments, an emerging hallmark of cancer \cite{pavlova2016emerging}. Indeed, it has been hypothesized that cancer cells may optimize their metabolic activity to ``go or grow,'' depending on local microenvironmental conditions \cite{DeDonatis2010}. Ordinarily, when ample oxygen is available, normal cells utilize mitochondrial oxidative phosphorylation and glycolysis in the cytoplasm to produce chemical energy in the form of ATP (Fig. \ref{fig:metabolism_background}A). Oxidative phosphorylation converts glucose to glycolysis intermediate products to pyruvate, which is oxidized in the mitochondria under aerobic conditions to produce 36 ATP (per glucose molecule) (Fig. \ref{fig:metabolism_background}A, i). In comparison, glycolysis occurs under anaerobic conditions to produce lactate and 2 ATP (per glucose molecule) (Fig. \ref{fig:metabolism_background}A, ii). There exists a metabolic trade-off whereby oxidative phosphorylation results in higher yield of ATP per glucose molecule with slower kinetics, while glycolysis results in lower yield of ATP per glucose molecule with considerably faster kinetics. Cancer cells often encounter hypoxic conditions, and may activate glycolysis via hypoxia-inducible factor 1$\alpha$ (HIF-1$\alpha$). This induces downstream transcription of glycolytic enzymes like hexokinase and lactate dehydrogenase \cite{semenza2013hif}, as well as enhanced expression of GLUT glucose transporters (Fig. \ref{fig:metabolism_background}A). Indeed, cancer cells benefit from using glycolysis to fulfil their metabolic requirements even when abundant oxygen is available. In particular, abundant ATP is advantageous for the nearly continuous cytoskeletal reorganization associated with directed cell migration \cite{dewane2021fueling}. Furthermore, enhanced lactate excretion via glycolysis by cancer cells acidifies the extracellular space (Fig. \ref{fig:metabolism_background}A, ii), which has been linked to the production of matrix metalloproteinases that facilitate migration through the ECM \cite{romero2016lactate}. Moreover, uncontrolled proliferation also requires glycolytic intermediates necessary for synthesis of nucleic acids, proteins, and lipids, as well as lactate which crucially maintains the NAD$^+$/NADH redox balance \cite{lunt2011aerobic}.  Cell metabolism, morphology, and migration are intimately linked, and thus targeting cell metabolism may present a novel strategy to inhibit the cellular energy production fueling collective cancer invasion.

J.S. Park et al investigated how the metabolism of individual human bronchial epithelial cells was modulated by soft or stiff planar substrates \cite{Park2020}. Morphologically, cells on stiff substrates exhibited larger footprints with extensive actomyosin stress fibers, while cells on soft substrates had smaller footprints with weak contractility. Consequently, cells on soft substrates showed reduced expression of all phosphofructokinase (PFK) enzymes, which catalyze a rate-limiting step of glycolysis by phosphorylating fructose 6-phosphate, and thereby determine the overall glycolytic rate (Fig. \ref{fig:metabolism_background}B, i). Loss of the PFK isoform PFKP on soft substrates could be attributed to degradation by the proteosome, specifically by the E3 ubiquitin ligase TRIM21. Indeed, TRIM21 associates with stress fibers and remains inactive when bound, thus maintaining PFKP activity. Remarkably, PFKP expression is elevated in patients with lung cancer, which could explain why tumors exhibit abnormally high levels of glycolysis. Genetic manipulation of these bronchial epithelial cells with a KRAS mutation resulted in consistently high PFK levels, which became insensitive to substrate stiffness. It should be noted that metabolic activity in breast cancer cell lines exhibit varying sensitivity when cultured over planar collagen hydrogels of varying density, thus modulating cytoskeletal activity (Fig. \ref{fig:metabolism_background}B, ii-iii) \cite{Mah2018}. Shear forces acting on cell-cell junctions can also increase glucose uptake, driving increased ATP production and glycolysis through AMPK activity (Fig. \ref{fig:metabolism_background}B, iv) \cite{bays2017linking}.

\begin{figure}[h!]
\centering
\includegraphics[width=3.5in]{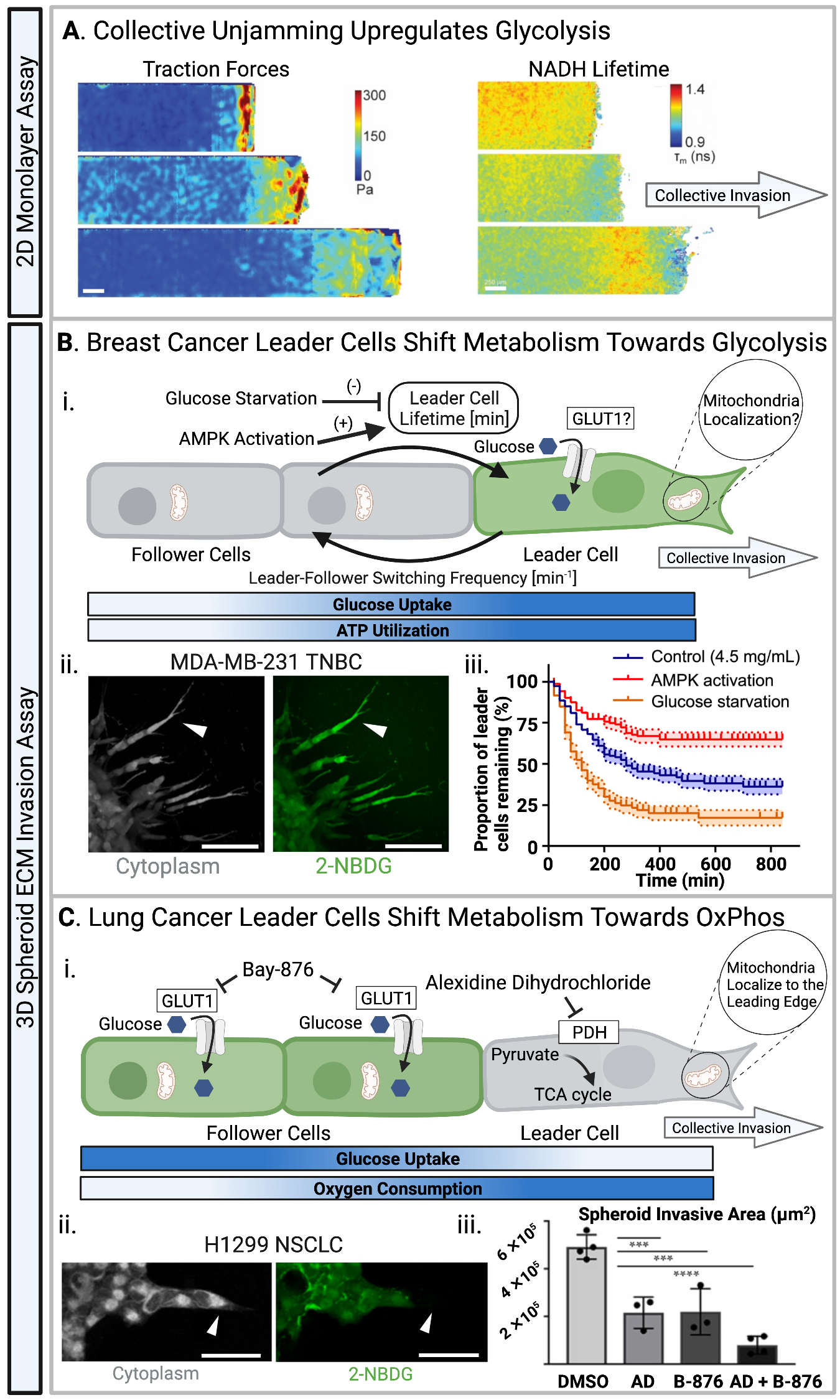}
\caption{A. Collective unjamming of epithelial monolayers reveals increase of traction forces and metabolic activity near invasion front. Reproduced from \cite{DeCamp2020}. B. Breast cancer leader cells exhibit increased glycolysis, but switch places regularly with followers. Reproduced from \cite{Zhang2019b}. C. Lung cancer leader cells exhibit increased oxidative phosphorylation, and invasion can be suppressed by targeting both pyruvate dehydrogenase and glucose uptake. Reproduced from \cite{Commander2020}.}
\label{fig:metabolism}       
\end{figure}

S.J. Decamp et al. showed that epithelial monolayers of canine kidney cells (e.g. MDCK II) exhibit spatially varying metabolic activity at collective fronts moving to occupy empty areas (Fig. \ref{fig:metabolism}A) \cite{DeCamp2020}. Indeed, collectively migrating cells at the front exhibited higher traction forces, elongated morphologies, and faster migration compared to cells towards the rear with small aspect ratios and low tractions. Moreover, cells at the collective front also display decreased NAD$^+$/NADH ratio, reduced NADH lifetime, and increased glucose uptake, which suggests a shift towards glycolysis. Interestingly, cells far to the rear also displayed a decrease in NADH lifetime and enhanced glucose uptake, indicating a shift towards glycolysis. However, these rearward cells showed an elevated cytoplasmic NAD$^+$/NADH ratio, which differs from cells at the collective front. This long distance signaling could be mediated through mechanical deformation of the compliant substrate, priming rearward cells for increased proliferation and migration. It should be noted that cell migration and proliferation into empty regions on top of planar substrates is relatively unimpeded and could differ significantly from metabolic behaviors in 3D.

Reinhart-King's group utilized a ratiometric fluorescent reporter of ATP: ADP ratio (PercivalHR) to elucidate how cellular energetics are regulated during individual and collective migration in 3D collagen matrix. First, highly metastatic breast adenocarcinoma cells (MDA-MB-231) expressing the PercivalHR construct invaded individually through narrow bifurcating channels in collagen I \cite{Zanotelli2019}. Cells exhibited greater glucose uptake and ATP:ADP ratio as they migrated through narrower channels, indicating that increased cell deformation was associated with higher energetic cost. Indeed, pharmacological treatment to perturb cytoskeletal stiffness and cell deformability was sufficient to modulate cell energetic state. Further, increasing matrix stiffness also increased energy costs, since cells required enhanced contractility to deform the matrix and enter confined spaces. When cells were presented with a choice of bifurcating into a wide or narrow channel, their decision-making was shown to be biased by the relative energetic cost of these two choices. Thus, cells were more likely to choose the narrower channel when the cells themselves or the matrix were more compliant.

In a complementary study, this lab also investigated how these same cells invaded collectively from multicellular spheroids, revealing that leader and follower cells invading in 3D matrix can dynamically switch their roles (Fig. \ref{fig:metabolism}B) \cite{Zhang2019b} \ref{fig:metabolism}B i).  Collective invasion was also impeded by denser 3D matrix with greater mechanical resistance, in qualitative agreement with the previous study using individual cells \cite{Zanotelli2019}. Nevertheless, collective invasion was associated with cooperative behaviors where leaders exhibited higher glucose uptake and energy consumption relative to followers. Once a leader cell exhausted its available ATP, it would switch with a follower, and this switching frequency increased in denser matrix. Moreover, expression of a fluorescent cell cycle reporter (e.g. CycleTrak) revealed that leader and follower cells exhibited behaviors consistent with the ``grow-or-go'' hypothesis \cite{DeDonatis2010}, whereby leader cells were less proliferative than follower cells, which was also observed using breast tumor organoids (e.g. MMTV-PyMT mice). 

In contrast, Commander et al. also observed metabolic heterogeneity in leader and follower cells invading from a different spheroid model, but concluded that followers exhibit higher glucose uptake than leaders (Fig. \ref{fig:metabolism}C ii) \cite{Commander2020}. Briefly, a non-small cell lung cancer line (H1299) was sorted for leaders and followers by photoswitching of a fluorescent label, followed by FACS \cite{Konen2017}. Interestingly, followers maintained distinct morphological and invasive phenotypes for several passages before reverting to the parental phenotype, while leader cells retained their phenotype indefinitely. Metabolic analysis based on extracellular flux (e.g. Seahorse Bioscience) revealed that follower cells typically exhibited increased glucose uptake, glycolysis and decreased oxidative phosphorylation \ref{fig:metabolism}C i). Instead, leader cells exhibited mitochondrial respiration and a pyruvate dehydrogenase dependency, which could be therapeutically targeted with alexidine dihydrocholoride \ref{fig:metabolism}C iii). A combination drug treatment against both pyruvate dehydrogenase and glucose uptake was effective at suppressing leader cell invasion and follower cell proliferation, respectively. It should be noted that this study utilized a different cell line (H1299) than Reinhart-King's investigations (MDA-MB-231), with varying degrees of phenotypic plasticity. Moreover, subtle differences in spheroid formation and matrix embedding procedures could result in very different phenotypic outcomes \cite{Peirsman2021}. 

An advantage of using engineered biomaterials is that cancer cells encounter highly consistent and homogeneous mechanical cues, so that the effect of stiffness or other matrix properties on metabolic activity can be systematically elucidated. However, it should be noted that cancer cells in vivo may exhibit very different metabolic states as they migrate through varying mechanical and biochemical microenvironments. Further, these cancer cells may interact dynamically with other stromal and immune cells, which could further act to enhance or suppress tumor progression. Thus, additional work is needed to understand how phenotypic plasticity of metabolic and migratory phenotypes is regulated in the context of mechanical cues from the surrounding microenvironment.

\section{The Extracellular Matrix and Engineered Biomaterials
}
\label{sec:ecm}

The extracellular matrix (ECM) is a complex structural network of proteins, glycoproteins, proteoglycans and polysaccharides that mechanically supports tissues while presenting instructive and permissive cues \cite{Cox2021}. ECM is continually being remodeled by a combination of protein deposition, post-translational chemical modifications, proteolytic degradation, as well as mechanically-driven reorganization \cite{Winkler2020}. These myriad processes are tightly regulated in normal tissues to maintain homeostasis, but become increasingly dysregulated during aging, fibrosis, and cancer progression \cite{Piersma2020}. Moreover, alterations in ECM structure, mechanics, and composition may regulate dormancy or metastatic colonization at distant tissue sites (i.e. the premetastatic niche) \cite{Peinado2017}. We briefly review relevant features of the ECM in the context of collective migration, and refer interested readers to more comprehensive reviews elsewhere \cite{Winkler2020,Cox2021,Piersma2020}. 

\begin{figure}[h!]
\includegraphics[width=4.5in]{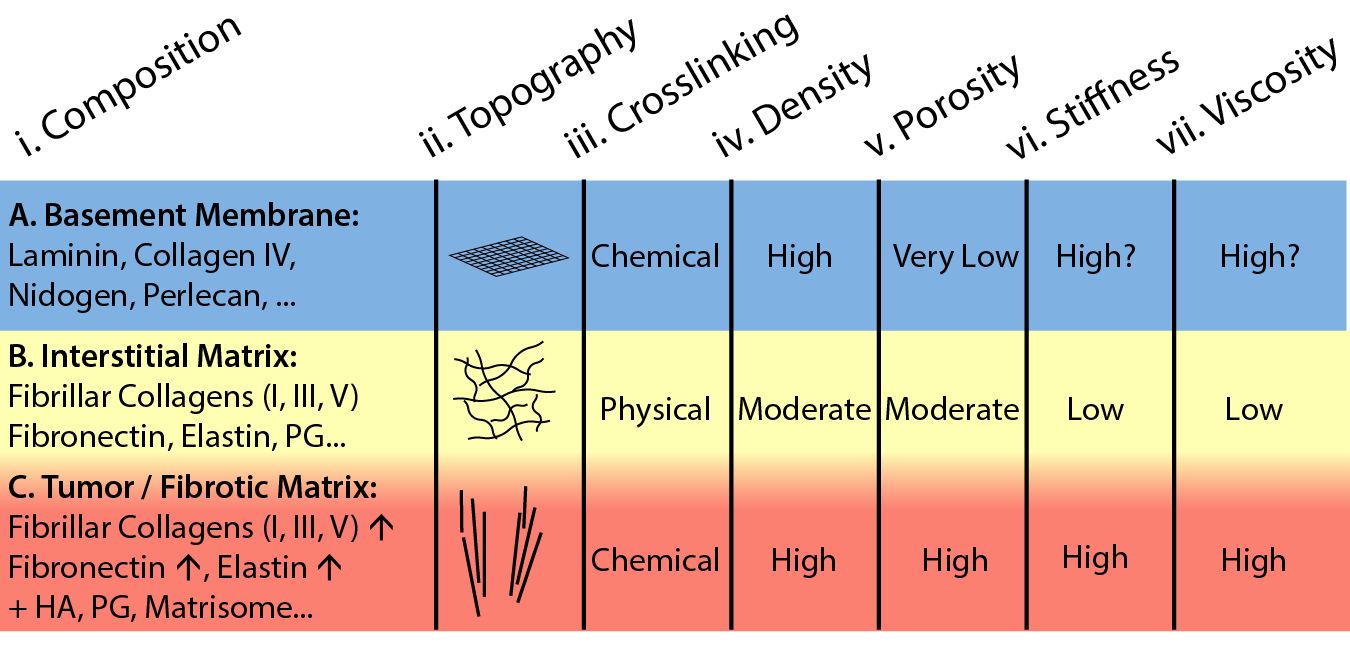}
\caption{Comparison of basement membrane (A), interstitial matrix (B) and tumor / fibrotic matrix (C) based on composition (i), topography (ii), crosslinking (iii), density (iv), porosity (v), stiffness (vi), and viscosity (vii). Abbreviations: HA, hyaluronic acid, PG, proteoglycans.}
\label{fig:ecmcartoon}       
\end{figure}

The basement membrane is a sheet-like structure that encloses epithelial tissues, comprised of laminin and collagen IV, linked through bridging proteins such as nidogen and perlecan (Fig. \ref{fig:ecmcartoon}A,i) \cite{Pozzi2017}. Indeed, laminin is more prominently displayed on the inner (epithelial-facing) side and instructs epithelial phenotype, while collagen IV may be more localized on the outer (stromal-facing) side to provide more structural support \cite{plodinec2015}. Altogether, these ECM proteins are assembled with a net-like architecture that is relatively thin (hundreds of nanometers) with small pores (tens to hundreds of nanometers) (Fig. \ref{fig:ecmcartoon}A,ii). Since the basement membrane is chemically crosslinked, high density, with very low porosity (Fig. \ref{fig:ecmcartoon}A,iii-v), it is not permissive for epithelial cell migration \cite{Kelley2014a}. Thus, enzymatic or mechanical remodeling of the basement membrane is crucial for carcinoma cells to invade into the surrounding stroma \cite{Chang2019a}.  Historically, reconstituted basement membrane (e.g. Matrigel) has been widely used as a compliant biomaterial that mimics the protein composition of native basement membrane \cite{Benton2014}. However, it should be noted that Matrigel is not of human origin, and is derived from mouse sarcoma cells (e.g. Engelbreth-Holm-Swarm) which exhibits appreciable batch-to-batch variability \cite{Aisenbrey2020}.

The interstitial matrix is a three-dimensional fibrous network comprised of fibrillar collagens (I, III, V), fibronectin, elastin, proteoglycans, glycosaminoglycans, and others (Fig. \ref{fig:ecmcartoon}B,i) \cite{Gritsenko2012}. In particular, collagen I is organized as randomly oriented fibers that are physically entangled at moderate density and porosity (Fig. \ref{fig:ecmcartoon}B,ii-v) \cite{Wolf2009}. Further, Keely and coworkers observed that these fibers exhibited a wavy (``crimped'') conformation \cite{Provenzano2006}, which has recently been reported to impede cell polarization \cite{Fischer2021} and enhance dormancy \cite{di2021tumor}. The interstitial matrix is occupied by stromal cells (e.g. fibroblasts), which play a crucial role in depositing and organizing ECM proteins \cite{Plikus2021}. It should be noted that the interstitial matrix can be spatially and mechanically non-uniform due to the presence of cell-sized spaces surrounded by denser matrix. Nevertheless, the typical pore structure of healthy interstitial matrix impedes cell migration without significant remodeling \cite{Wolf2009}. A biomimetic model for these fibrous architectures is to reconstitute collagen I from rat tail or bovine skin, which can be tuned based on collagen concentration, polymerization temperature and pH (see reviews in \cite{Hapach2015,Sapudom2018}). 

Pathological states such as fibrosis and tumor progression are associated with aberrant deposition of collagens, fibronectin, elastin, laminin, hyaluronic acid, proteoglycans (e.g. versican, syndecan, glypican, etc.), and other matrisome proteins (e.g. tenascin C, periostin, osteopontin, SPARC, thrombospondin) (Fig. \ref{fig:ecmcartoon}C,i) \cite{Piersma2020}. Structurally, tumor-associated collagen I is increasingly dense, straight and aligned, with cell-sized tracks leading into the stroma (Fig. \ref{fig:ecmcartoon}C,ii-v) \cite{Provenzano2006,Provenzano2008b,Conklin2011}. Indeed, collagen I can be further strengthened by chemical crosslinking by lysyl oxidase and transglutaminase (Fig. \ref{fig:ecmcartoon}C,iii) \cite{Levental2009}, which has been recently linked to the inflammatory activity of tumor-associated macrophages \cite{Maller2021}.

The stiffness of tumor associated ECM is dramatically higher than healthy ECM, an indicator of aberrant ECM composition and architecture (Fig. \ref{fig:ecmcartoon}B,vi;C,vi) \cite{Piersma2020}. For any solid material, these mechanical properties can be quantified based on a reversible deformation in response to an applied stress (i.e. elasticity), which is typically linear for small deformations (i.e. strains) \cite{fung_tong_2001}. Interestingly, fibrous materials often exhibit a nonlinear elasticity, so that they can stiffen at increasing strain, attributed to the increasing difficulty of reorienting fibers, or the cost of straightening and stretching fibers \cite{Broedersz2014}. For sufficiently large deformations, this structural reorientation may be permanent (i.e. plastic) and will not be reversible after the stress is removed. Finally, it should be noted that biomaterials incorporate significant (viscous) fluid, resulting in a time-dependent mechanical response as some energy is dissipated via viscoelastic or poroelastic mechanisms \cite{Chaudhuri2020}. An unresolved question is the stiffness and viscosity of the basement membrane, which varies across tissues and also is challenging to probe due to its small thickness and tight integration with surrounding tissue (Fig. \ref{fig:ecmcartoon}A,vi-vii). Interestingly, H. Li et al. recently reported that cell-deposited basement membranes exhibit nonlinear strain stiffening, which has not been observed using other mechanical measurements that operate in the linear regime \cite{Li2021}.

Synthetic hydrogels can be prepared with comparable stiffness as ECM using flexible polymers with tunable crosslinks and ligand density, resulting in spatially uniform materials with nanoscale pores \cite{Leggett2017,Beri2018}. For example, polyethylene glycol and polyacrylamide have been widely explored to elucidate how cancer cells respond in vitro to linear elastic material properties (see review in \cite{Janmey2020}). In particular, polyethylene glycol permits systematic control of ligand density, crosslinking and degradability to investigate cancer cell invasion \cite{gill2012synthetic, beck2013independent, singh2015synthetic,pradhan2019tunable}. Alternatively, alginate is a seaweed-derived polysaccharide where viscoelasticity (e.g. stress relaxation) can be tuned using reversible ionic crosslinks or by molecular weight \cite{Chaudhuri2017}. Furthermore, synthetic hydrogels with spatially anisotropic architectures are able to mimic ECM cues, such as the aligned topographical features of collagen fibers \cite{Prince2019}.  Such ``designer biomaterials'' may permit improved control over distinct mechanical and biochemical features of the cell microenvironment. For example, aligned ECM cues can be patterned within a homogeneous matrix using magnetic nanoparticle self-assembly, allowing for hydrogels with modulated alignment cues while maintaining comparable elastic moduli and ligand densities \cite{kim2016independent,paul2019probing}.

\section{Precision Measurement of \textbf{Cell and} ECM Mechanics
}
\label{sec:mechanics}

Invading cancer cells both sense and respond to the structure and composition of the ECM, a bi-directional interaction known as dynamic reciprocity \cite{Xu2009e}. Fluorescence imaging of cell morphology is widely used to elucidate how cells are shaped by the local ECM properties \cite{Driscoll2015a}. However, visualizing how cells act on the surrounding ECM remains challenging, particularly in 3D geometries \cite{Polacheck2016}. Seminal work by Legant et al. visualized how fibroblasts applied mechanical tractions within a synthetic 3D hydrogel (e.g. polyethylene glycol diacrylate, PEGDA) by tracking the motion of microscale tracer particles relative to an undeformed reference state (e.g. traction force microscopy) \cite{Legant2010}. It should be noted that PEGDA is a purely elastic material, with a well-established constitutive equation that permits the cell-generated stresses to be quantitatively extracted from the matrix strain field. Nevertheless, naturally-derived biomaterials exhibit a more complex rheological response that is nonlinear and viscoelastic, and can be irreversibly remodeled by cell secreted factors \cite{Hall2013}.

Both J. Steinwachs et al. and M. Hall et al. measured tractions exerted by single breast cancer cells invading within reconstituted collagen I matrix labeled with fluorescent microparticles \cite{Steinwachs2016,Hall2016}. These breast cancer cells (e.g. MDA-MB-231) mechanically interact with discrete collagen fibers, which become aligned along the cell axis to locally stiffen the matrix. This nonlinear strain-stiffening response suggests that cells within these fibrillar materials can mechanically interact over longer distances than in more homogeneous materials that exhibit a linear elastic response. In order to compute cell-generated stresses, both groups formulated constitutive equations optimized for random fiber networks, validated using bulk rheology of cell-free collagen I matrix. Interestingly, recent work by van Oosten et al. shows that these strain-stiffening fiber networks switch to strain-\emph{softening} behavior in the presence of cell-sized inclusions, reminiscent of the architecture and mechanics of living tissues \cite{vanOosten2019}. Thus, non-uniformity of the ECM at multiple length scales is likely to impact cellular mechanobiology.

\begin{figure}[h]
\centering
\includegraphics[width=4in]{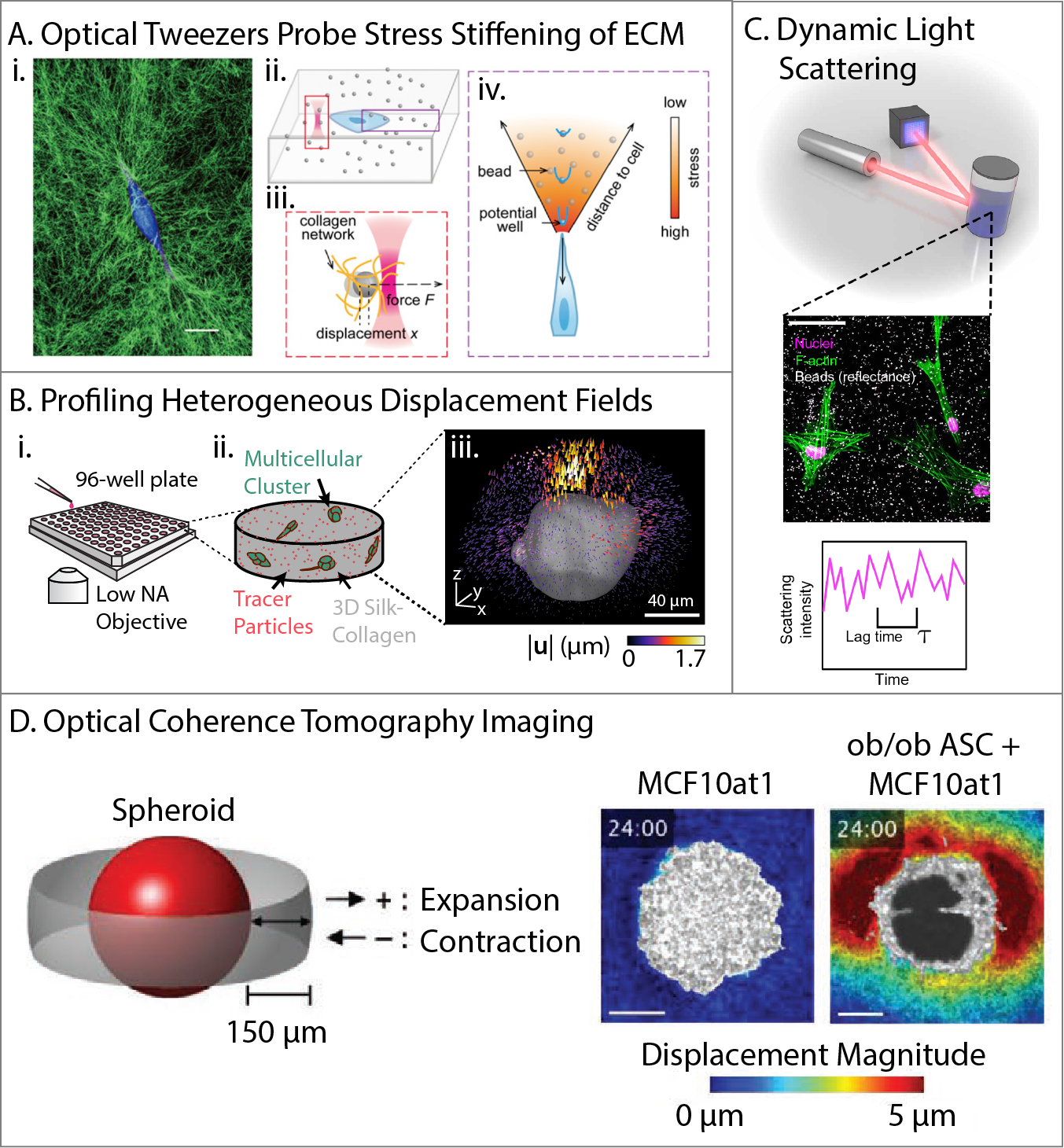}
\caption{A. Direct measurement of ECM stress stiffening using optical tweezers. (i) MDA-MB-231 cell (blue) in a 3D collagen network (green). (Scale bar, 10 µm.) (ii-iv) Schematics of the force–displacement measurement using optical tweezers, and the relation between matrix stiffening (blue potential wells) and the cell-generated stress field along the cell axis. Reproduced from \cite{Han2018}. B. High-throughput confocal imaging on a 96 well plate (i), visualizing multicellular clusters (green) and tracer particles (red) embedded in 3D silk collagen matrix (ii). Representative displacement field exhibits spatially non-uniform regions of local outward (yellow) and inward (purple) displacement (iii). Reproduced from \cite{Leggett2020}. C. Dynamic light scattering based on cells encapsulated within 3D collagen / Matrigel, read out via fluctuations in scattering intensity over varying lag times. Reproduced from \cite{Krajina2021}. D. Optical coherence tomography of multicellular spheroids in 3D collagen matrix for label-free readout of matrix deformation, revealing enhanced invasion for mammary epithelial cells (MCF-10A) co-cultured with obese adipose stem cells (ob/ob ASC). Reproduced from \cite{Ling2020}.}
\label{fig:ecm}       
\end{figure}

Y. Han et al. sought to directly measure matrix mechanical properties in the neighborhood of single cells using optical tweezers \cite{Han2018} (Fig. \ref{fig:ecm}A,i,ii). In these experiments, microparticles entangled within a collagen I matrix were perturbed under a controlled displacement $x$ and the resulting force $F$ was optically measured, which is a readout of the local matrix stiffness (Fig. \ref{fig:ecm}A,iii). This approach mapped the spatial variation in stiffness to a nonlinear stress profile Fig. \ref{fig:ecm}A,iv). Notably, matrix stiffness was strongly elevated near cancer cells (e.g. MDA-MB-231) as the fibrillar matrix was plastically deformed into more aligned architectures, which was also observed for reconstituted basement membrane and fibrin. Although this method is a very sensitive and localized measurement of matrix rheology, it is relatively time and labor intensive since microparticles are probed one at a time.

Multicellular clusters in 3D matrix may coordinate their tractions to facilitate collective migration, which adds additional complexity to be explored. S. Leggett et al. investigated how mammary epithelial acini (e.g. MCF-10A) transitioned towards invasion after controlled induction of EMT via the master regulator Snail \cite{Leggett2020}. Cells were embedded within composite silk-collagen matrix \cite{Khoo2019} in a 96-well plate, enabling increased experimental throughput to evaluate drug response (Fig. \ref{fig:ecm}B). Using optimized topology-based particle tracking, localized patterns of protrusive and contractile matrix displacement could be resolved with submicron resolution \cite{Patel2018}. Notably, epithelial clusters were usually compact and exhibited several regions of local protrusion or contraction. Induction of a transitory EMT state results in clusters with extended protrusions with more regions of local contraction, indicative of outgrowth and invasion. Finally, fully mesenchymal states were associated with spindle-like morphologies with only a few localized contractile regions and minimal protrusions. These results show that collective migration and EMT are associated with spatially non-uniform patterns of matrix displacement, which can be used as a mechanophenotypic ``signature'' of cell state. This platform is scalable for a broader range of cell types and biomaterials, which is particularly relevant or clinical translation using patient-derived cells and drug screening.

In order to longitudinally probe matrix viscoelasticity over a wide frequency range, B. Krajina et al. utilized dynamic light scattering to measure fluctuations of dilute microparticles within a biomaterial (Fig. \ref{fig:ecm}C) \cite{Krajina2021}. Based on correlations in scattering intensity, an ensemble averaged mean square displacement can be extracted and converted to a frequency-dependent shear modulus using the generalized Stokes-Einstein relation \cite{Squires2010}. In principle, this approach could resolve both thermally driven motion as well as actively driven motion due to cell-generated forces. For instance, mammary epithelial spheroids (e.g. MCF-10A) embedded in composites of collagen I and Matrigel drove considerable changes in shear modulus over 6 days. With the addition of TGF-$\beta$ to drive invasion and EMT, there was an effective increase in matrix stiffness at high frequencies and matrix fluidization at low frequencies, which was explained based on elevated matrix degradation as well as cellular contractility. This dynamic light-scattering approach promises to provide new insights into how biomaterials are altered over extended durations in response to cellular activity, although it will likely require complementary measurements with conventional fluorescence microscopy to map back to cell morphology and migratory behaviors. 

Finally, Adie and coworkers utilized optical coherence tomography for label-free imaging of cells and collagen fibers \cite{Mulligan2019}. As a case study, mammary epithelial spheroids (e.g. MCF-10A) were characterized with or without co-culture with adipose stem cells \cite{Ling2020}. Notably, mammary epithelial spheroids only remained relatively localized with limited matrix deformation (Fig. \ref{fig:ecm}D). In comparison, the addition of adipose stem cells from obese mice resulted in extensive collective invasion as well as matrix remodeling, which could be rescued by pharmacological inhibition of matrix degradation or Rho kinase activity. This technique has great potential since it can be scaled to larger imaging volumes in highly scattering media, directly imaging biomaterial deformations with minimal phototoxicity.

Emerging technologies to visualize bi-directional interactions between cancer cells and 3D matrix will enable new insights into collective invasion and drug response. It should be noted that cell mechanics are challenging to measure deep inside 3D matrix, since most characterization techniques require direct contact with the cell (e.g. micropipette aspiration, atomic force microscopy, etc.). Nevertheless, techniques such as particle tracking microrheology or optical tweezers can directly probe intracellular rheology in 3D, as we have recently reviewed elsewhere \cite{yiwei2022}. We also envision future explorations that utilize patient-derived organoids, as well as stromal and immune cells. Thus far, these early proof-of-concepts have investigated cell-generated tractions in relatively simplified biomaterials such as collagen I or reconstituted basement membrane (e.g. Matrigel). As the composition and architecture of biomaterials increases in complexity to mimic the in vivo environment, an improved fundamental understanding of these material properties will be needed to understand how cancer cells are affected. Lastly, these biophysical measurements all require specialized equipment and training for optimal results, and further work is needed to translate these techniques into conventional biomedical settings.

\section{Epithelial-Mesenchymal Plasticity and Collective Cell Migration
}
\label{sec:EMT}
Classically, EMT has been understood as a phenotypic switch in which adherent cells lose epithelial biomarkers (e.g. E-cadherin) and gain mesenchymal biomarkers (e.g. vimentin) in order to disseminate individually for embryonic development and wound healing \cite{Nieto2016}. EMT is often observed at tumor invasion fronts, mediated by tumor-stromal interactions and the interstitial matrix \cite{christofori2006new}. More recently, epithelial-mesenchymal plasticity (EMP) has been proposed to represent a broader spectrum of intermediate or hybrid cell states that may exhibit some mixture of epithelial and mesenchymal features \cite{Yang2020b}. Functionally, EMP is associated with a decrease in differentiated features (e.g. apico-basal polarity, strong cell-cell adhesions), resulting in reorganization of the cytoskeleton and cell-matrix adhesions towards front-back polarity and invasion across the basement membrane, which may or may not be captured by certain biomarkers or classical EMT transcription factors (e.g. Snail, Twist, ZEB, etc.) \cite{Yang2020b}. These morphological phenotypes often exhibit appreciable variability at the single cell level, even for relatively controlled induction of transcription factors \cite{leggett2016morphological}. Nevertheless, the mechanistic role of EMT in human tumor metastasis remains unresolved, partially due to the difficulty of measuring a rare and dynamic process at the single cell level (within patients) \cite{Diepenbruck2016}. For instance, if EMT occurs transiently as tumor cells disseminate from the primary site and they revert back to an epithelial state (MET) at a secondary metastatic site, measurements of the primary and metastatic site will most likely yield epithelial cells \cite{Brabletz2012}. Several recent papers have combined live cell imaging with deep molecular profiling in order to better elucidate the potential role of EMP.

\begin{figure}[h]
\centering
\includegraphics[width=3in]{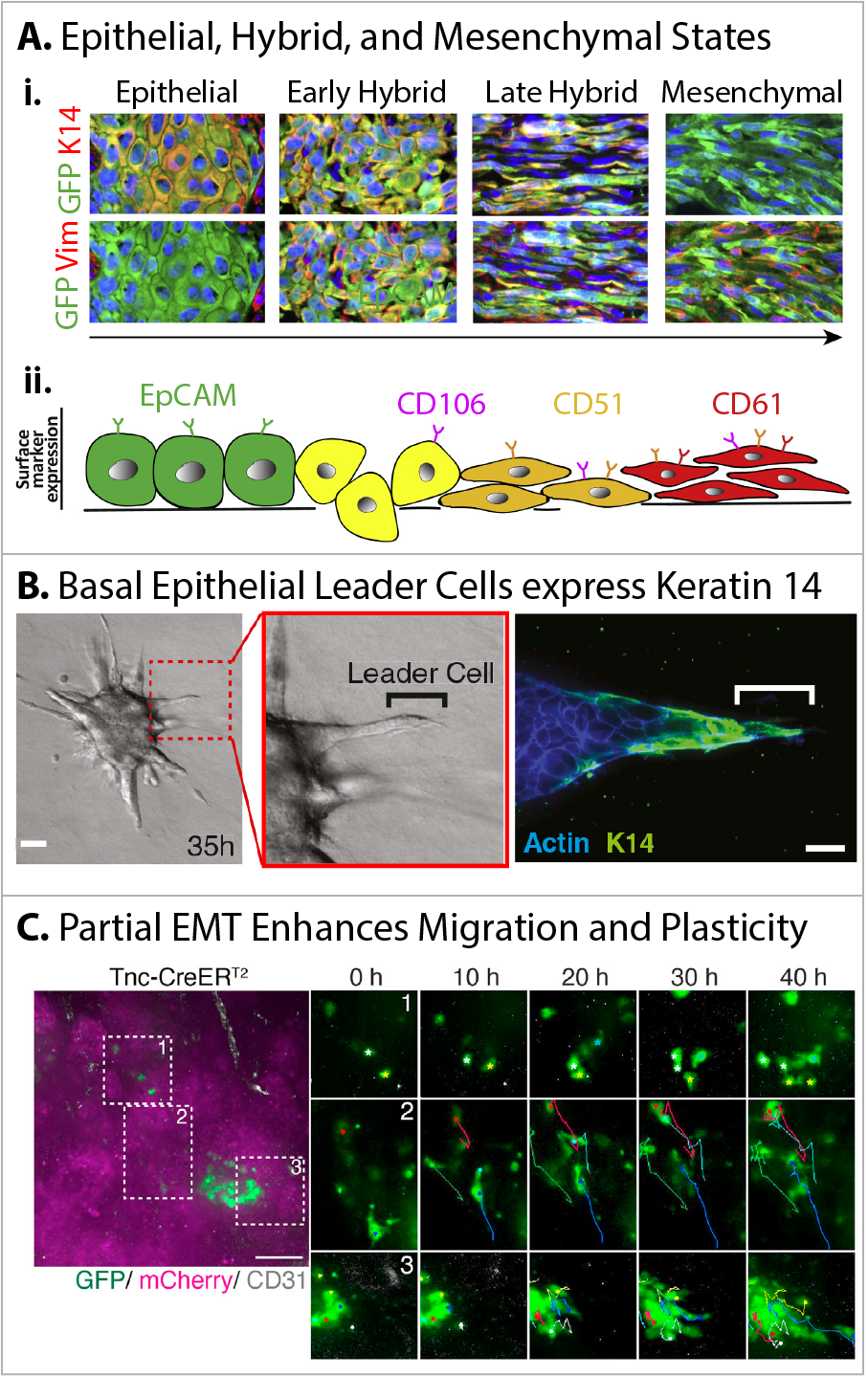}
\caption{A. Immunofluorescence staining of keratin 14 (K14) and vimentin (Vim) (red) expression in tumor cells (GFP) across epithelial, early/late hybrid and mesenchymal states (i), corresponding to changes in EpCAM, CD106/Vcam1, CD51/Itgav, and CD61/Itgb3 (ii). Reproduced from \cite{Pastushenko2018,Pastushenko2019}. B. Leader cells in MMTV-PyMT breast tumor organoids exhibit a basal epithelial phenotype with high keratin 14 expression. Reproduced from \cite{Cheung2013}. C. Lineage tracing of partial EMT via Tenascin C gene expression reveals enhanced migration and plasticity relative to late EMT. Reproduced from \cite{Luond2021}}.
\label{fig:emt}       
\end{figure}

I. Pastushenko et al. utilized fluorescence activated cell sorting (FACS) and single cell RNA-seq to map an EMT landscape in genetically engineered mouse models of skin squamous cell carcinoma (KRas$^{LSL-G12D}$p53$^{fl/fl}$) and breast cancer (MMTV-PyMT) that overexpressed fluorescent proteins \cite{Pastushenko2018, Pastushenko2019}. They defined an epithelial state based on high expression of Keratin 14 (K14) with no vimentin expression, which exhibited a compact cell morphology and (high E-cadherin expression) (Fig. \ref{fig:emt}A,i). Further, hybrid states were observed which coexpressed K14 and vimentin with partially elongated morphology (and low E-cadherin expression), associated with collective migration. Finally, mesenchymal states expressed vimentin with no K14 (or E-cadherin), and exhibited a fully elongated morphology. The epithelial state also expressed high EpCAM, which was absent in hybrid and mesenchymal states (Fig. \ref{fig:emt}A,ii). Instead, an ``early hybrid'' state was defined based on the absence of CD106 (Vcam1), CD51 (Itgav), and CD61 (Itgb3), whereas a ``hybrid state'' was associated with CD106 expression only. Next, a ``late hybrid'' state was associated with expression of CD51 and/or CD106. A mesenchymal state was associated with co-expression of CD51 and CD61 or CD51, CD106, and CD61. Functionally, epithelial cells proliferated faster than cells in hybrid or mesenchymal states. However, hybrid or mesenchymal cells exhibited much higher tumor initiating potential than epithelial cells when implanted at successively smaller dilutions and displayed increased invasion. Nevertheless, hybrid cells exhibited the most plasticity to transition to other states, as well as the greatest metastatic potential. These hybrid or mesenchymal states were also promoted by the local microenvironment, including blood and lymphatic vessels as well as macrophages.

Instead, K.J. Cheung et al demonstrated that epithelial cells expressing K14 from genetically engineered mouse models of breast cancer (MMTV-PyMT) act as leader cells for collective migration (Fig. \ref{fig:emt}B) \cite{Cheung2013}. Mouse tumors were isolated and digested into fragments, which were then embedded in 3D collagen I matrix. Leader cells were associated with basal markers (e.g. p63, P-cadherin, and Keratin 5), but did not express common EMT biomarkers (e.g. vimentin, Snail, Twist). Further, these leader cells maintained E-cadherin expression and were mechanically connected to their followers, which were K14 negative. Indeed, K14 positive leader cells were preferentially localized at the tumor periphery and in lung metastases in mouse models, as well as in organoids and histology slides from human patients. Interestingly, K14 positive cells were present but did not function as leaders when these organoids were cultured in Matrigel. Re-transplantation of these organoids from Matrigel to collagen I rescued the K14 positive leader cell phenotype and aggressive invasion. 

F. L{\"u}{\"o}nd et al. developed a dual recombinase fluorescence reporter construct to lineage trace breast tumor cells that had undergone an EMT (reversible or irreversible) \cite{Luond2021}. Briefly, a first reporter labeled epithelial cells with mCherry, but irreversibly switched to GFP expression when early EMT (e.g. Tenascin C, Tnc) was induced with Tamoxifen treatment. A second reporter similarly labeled epithelial cells, but irreversibly switched to GFP expression during late EMT (e.g. N-cadherin, Cdh2), also under Tamoxifen treatment. Live cell imaging of primary tumor slices revealed that some early (partial) EMT cells exhibited an elongated morphology as individuals with directional migration (Fig. \ref{fig:emt}C,1,2). Other early (partial) EMT cells associated with multicellular ``colonies,'' which maintained cell-cell contacts, although cells at the periphery of these colonies were elongated and motile (Fig. \ref{fig:emt}C,3). In comparison, late (full) EMT cells were also elongated but remained localized in perivascular regions near capillaries. Moreover, early (partial) EMT cells invaded collectively and aggressively as leader cells when seeded on collagen I matrix, while late (full) EMT cells invaded individually. Further, early (partial) EMT cells were enriched in lung metastases, while late (full) EMT cells were enriched in response to chemotherapy. Other recent reports corroborate that partial EMT is associated with collective invasion and plasticity \cite{Aiello2018}, as well as enhanced metastatic potential \cite{Simeonov2021}.

\section{Perspective and Future Directions
}
\label{sec:perspective}
The statistician G.E.P. Box famously commented that ``\emph{all models are wrong; the practical question is how wrong do they have to be to not be useful}.'' \cite{boxmodel}
This aphorism is relevant in the context of cancer biology, where modeling refers to vitro or in vivo experimental systems that represent key features of patient tumors \cite{Ben-David2019}, as well as the physical sciences, where modeling refers to theoretical and computational approaches \cite{Altrock2015}. In both instances, an unresolved question is how a group of cancer cells with significant (intratumoral) heterogeneity are able to coordinate robust behaviors \cite{Tabassum2015}. As presented here, physical science approaches offer promising tools to elucidate generalized principles of collective invasion, given that cancer has common set of requirements for metastasis despite high intratumoral heterogeneity. We argue that such collective processes are crucial for invasion and metastasis as well as therapy resistance, which are ultimately responsible for most cancer-related fatalities in human patients \cite{Weiss2022}. Thus, in this chapter, we have considered new biological and engineering approaches that have the potential for useful insights, given the ethical and practical limits on direct experimental access to human patient tumors. 

Modern single cell -omics enable unprecedented insights into molecular signaling pathways that regulate cellular behavior \cite{Satija2019}. However, such deep phenotyping typically requires the removal and destruction of cells at an endpoint, making it difficult for dynamic measurements. In comparison, live cell imaging and fluorescent reporters can visualize how cells migrate and proliferate with unprecedented spatiotemporal resolution \cite{Lemon2020,Specht2017}. However, these imaging techniques are limited to a few fluorescence channels, so that only a few signaling pathways or cellular features can be monitored. This mismatch makes it challenging to elucidate how myriad molecular signaling pathways are integrated into mechanobiological phenotypes such as cell migration and proliferation. The Connectivity Map approach has proven quite powerful to computationally link gene expression profiles to drug response \cite{lambcmap}. An intriguing possibility is to further couple such molecular analyses with cellular-scale computational models that account for the mechanobiology of the cytoskeleton, cell-matrix adhesions, etc. \cite{Buttenschon2020} This could make testable predictions about therapeutic regimens that could inhibit invasion or proliferation. Given interpatient heterogeneity and the challenges of drug development, even predicting which treatments may \emph{not} work would be useful to inform clinical practice. 

Aberrant metabolism represents a promising target to inhibit both collective migration as well as uncontrolled proliferation. Leader cells represent an intriguing target, either via inhibition of glycolysis (e.g. 2-Deoxyglucose) or oxidative phosphorylation (e.g. metformin). Given the observed phenotypic plasticity and heterogeneity of leader and follower cells, it may be necessary to target both glycolysis and oxidative phosphorylation. It should also be noted that various FDA-approved drugs that inhibit metabolism can in turn mitigate aggressive EMT features of tumors \cite{ramesh2020targeting}. For example, when activated by excess glucose, the polyol metabolic pathway can induce EMT through autocrine TGF-$\beta$ stimulation \cite{Schwab2018}, which is known to potentiate tumor aggressiveness. The polyol pathway inhibitor epalrestat, used in the treatment of diabetic neuropathy, is under a Phase II clinical trial to evaluate its potential to inhibit triple-negative breast cancer \cite{Wu2017a}. Thus, an improved fundamental understanding of how metabolism and mechanobiology are mechanistically coupled may inform the design of targeted therapies, likely acting in combination on different hallmarks of cancer.

Although visualizing cell morphology and migration with fluorescence microscopy is relatively straightforward, inferring cell-cell interactions remains challenging, and is often indirectly inferred from cell morphology and migration. For instance, physics-inspired analyses based on PIV or order parameters are sufficient to reveal when groups of cells migrate with similar speed and direction \cite{Ilina2020}, but provide limited quantitative information on how these cells interact. One promising approach for analyzing multicellular configurations is to analyze the mechanical cost of rearrangements relative to some idealized reference state \cite{Yang2021}. Alternatively, the topological differences between different multicellular architectures can be compared using persistence homology \cite{Bhaskar2021}. It would be intriguing to develop a hybrid machine learning approach that accounts for the mechanical cost of cell reorganization but does not assume a reference state. However, it should be noted that tumor cells are intrinsically heterogeneous, and approximating them as identical agents with exactly the same cell-cell and cell-matrix adhesions may not be valid. Establishing meaningful readouts of heterogeneity and cooperative behavior may nevertheless be useful as a therapeutic target (beyond cell viability) that can be perturbed. Moreover, understanding when groups of cells exhibit more homogeneous behaviors may provide insight into how ``noise'' is regulated by homeostasis in normal epithelial tissues.

Similarly, elucidating reciprocal mechanical interactions between cells and ECM is experimentally difficult \cite{Polacheck2016}. Early groundbreaking work utilized planar 2D substrates with controlled biochemical ligand density and stiffness, which were non-cell-degradable and exhibited a well-defined linear elastic response \cite{Janmey2020}. Subsequent experiments using 3D biomaterials are based on a similar premise - that cells should exhibit relatively similar behaviors when cultured within (relatively) reproducible and spatially homogeneous mechanical environments \cite{LeSavage2021}. Nevertheless, cells also exhibit the capability to substantially remodel biomaterials, so that the local architecture and rheology may vary considerably at later times. Thus, recent technology development seeks to mechanically probe alterations of ECM \cite{Han2018,Mulligan2019,Krajina2021}, although there remain trade-offs in spatial and temporal resolution. Alternatively, spatially heterogeneous patterns of forces may be sufficient to infer distinct cell states \cite{Leggett2020}. Thus, we envision that dynamic, stimuli-responsive biomaterials will be increasingly used to reveal dynamic cell responses \cite{Uto2017}. Moreover, it should be noted that ECM in vivo is spatially \emph{in}homogeneous in composition and structure, particularly during fibrosis or tumor progression. Highly tunable biomaterials capable of recapitulating desmoplastic ECM architectures in the tumor microenvironment \cite{Prince2019} will be needed to understand why current therapies, such as CAR-T immunotherapy, often fail to penetrate immunosuppressive solid tumors \cite{Pires2021}. High resolution mapping of local ECM architecture in vivo, combined with a physical understanding of how cells migrate collectively in different homogeneous biomaterials in vitro, could also predict how cell might invade inhomogeneous microenvironments in vivo \cite{Madsen2015}. Indeed, robust biomaterial-based models serve as an important first-step in elucidating collective invasion mechanisms in a controlled environment relative to in vivo models, where mechanisms can be further corroborated. 

Finally, tumor cells not only exhibit phenotypic heterogeneity but also phenotypic plasticity, which is often a relatively rare event. Thus, it remains challenging to test the role of epithelial-mesenchymal plasticity in human patients without complete single cell information at high temporal resolution \cite{Yang2020b}. Indeed, it is conceivable that tumor cells could adapt their phenotype as they encounter different microregions of the primary tumor, colonize the secondary metastatic site, or in response to therapeutic treatments \cite{gupta2006cancer}. Recent advances in biofabrication and microfluidics enable compartmentalized engineered tissues, where soluble signaling conditions can be rapidly switched \cite{Ayuso2021a}. This capability to rapidly and sharply switch microenvironmental cues is likely to be quite powerful to directly observe how cells respond. Further, improvements in epigenetic profiling will be useful to understand when cells mechanically adapt or retain a ``memory'' of prior phenotypes  \cite{Nemec2021}. An ongoing translational challenge may be to understand how patient tumor cells individually and collectively adapt to changing microenvironmental conditions via leader cells or  EMP, and to limit heterogeneity in an evolutionary context \cite{Gatenby2020}.

In conclusion, we have considered how the mechanobiology of cell-cell and cell-matrix interactions mediate how tumor cells collectively invade into 3D matrix. We highlight how these coordinated behaviors with leader-follower interactions and epithelial-mesenchymal plasticity represent a caricature of embryonic development, as tumor cells repurpose the associated molecular signaling and biophysical machinery. We also highlight emerging technologies in ex vivo measurements of cell migration based on 3D biomaterials with live cell imaging. In combination, these technologies can quantify how groups of cells cooperate mechanically or metabolically with each other and to remodel the surrounding ECM. These exquisitely sensitive measurements of single cell dynamics will enable new physical models that could be integrated with molecular profiling to predict how patient cells respond to targeted treatments, revealing new fundamental insights and accelerating clinical translation. Ultimately, we envision that the integration of new bioengineering assays with physical modeling represent a highly ``useful'' representation of human tumors to address unresolved questions of intratumoral and interpatient heterogeneity in cancer progression.

\section*{Acknowledgements}
We apologize to authors whose work could not be included due to space constraints. This work was supported by the National Institute of General Medical Sciences (R01GM140108, P20GM109035).

\bibliography{references}

\end{document}